\documentclass{ws-p10x7}
\newcommand{\GeV}{\rm GeV}
\newcommand{\an}{\overline{\alpha}_0}
\newcommand{\ao}{\overline{\alpha}_1}

\newcommand{\as}{\alpha_s}
\newcommand{\asmz}{\alpha_s(M_Z)}
\newcommand{\mean}[1]{\left< #1 \right>}

\newcommand{\mi}{\mu_{\scriptscriptstyle I}}

\begin{document}

\title{Event Shape Studies at HERA}

\author{Hans-Ulrich Martyn}

\address{I. Physikalisches Institut der RWTH, D--52056 Aachen, Germany \\
         (on behalf of the H1 and ZEUS collaborations) }

\twocolumn[\maketitle
  \abstract{ 
    Recent progress on the study of power corrections applied to 
    event shape variables in deep inelastic $e p$ scattering is discussed.}]

\section{Introduction}

Event shape variables in deep inelastic scattering allow 
to study QCD properties of the final state.
Recent results based on high statistics data by the {\sc Zeus}\cite{zeus}
and  H1\cite{h1} collaborations at {\sc Hera} are discussed
with emphasis on power law hadronisation effects.

A suitable frame of reference 
with optimal separation between the current jet 
and the proton remnant is the Breit system, where the spacelike
$\gamma/Z$ with momentum $Q$ 
collides head-on with an incoming quark of momentum $Q/2$.
The quark is back-scattered into the current hemisphere, 
while the proton fragments into the remnant hemisphere
(QPM picture).
Event shapes studied in the current region\cite{zeus,h1} are: 
thrust $\tau$ and $\tau_c$, broadening $B$, $C$ parameter and 
jet mass $\rho$ and $\rho_0$. 
H1 also investigates two-jet rates, 
the transitions from  $(2+1)$ to $(1+1)$ jets,
in the whole phase space:
$y_{fJ}$ for a factorisable {\sc Jade} and $y_{kt}$ for the $k_t$ jet algorithm.
The analyses of both experiments are very similar. 
The kinematic range covers $Q = 7 - 140~\GeV$, {\sc Zeus} distinguishes
two  $x$-bins at low $Q < 18~\GeV$.
The data are unfolded to the hadron level, where H1 takes the true masses
and {\sc Zeus} assumes massless hadrons. 
This  difference \mbox{affects} jet masses and two-jet rates.

\section{Power corrections to mean values of event shapes}

The mean value of an event shape variable $F$ can be expressed as\cite{dw}
\begin{eqnarray}
  \langle F \rangle & = &
  \langle F \rangle^{{\rm pert}} + a_F  {\cal P} \, .
  \label{eq1}
\end{eqnarray}
The perturbative part $\langle F \rangle^{{\rm pert}}$
is calculated in ${\cal O}(\alpha^2_s)$  using {\sc Disent}\cite{catani}.
The program {\sc Disaster++}\cite{graudenz} yields consistent but somewhat higher
mean values\cite{h1}, at most a few per cent for $\langle B \rangle$ at low Q.
These discrepancies have no influence on the conclusions. 

\begin{figure} \centering
  \epsfig{file=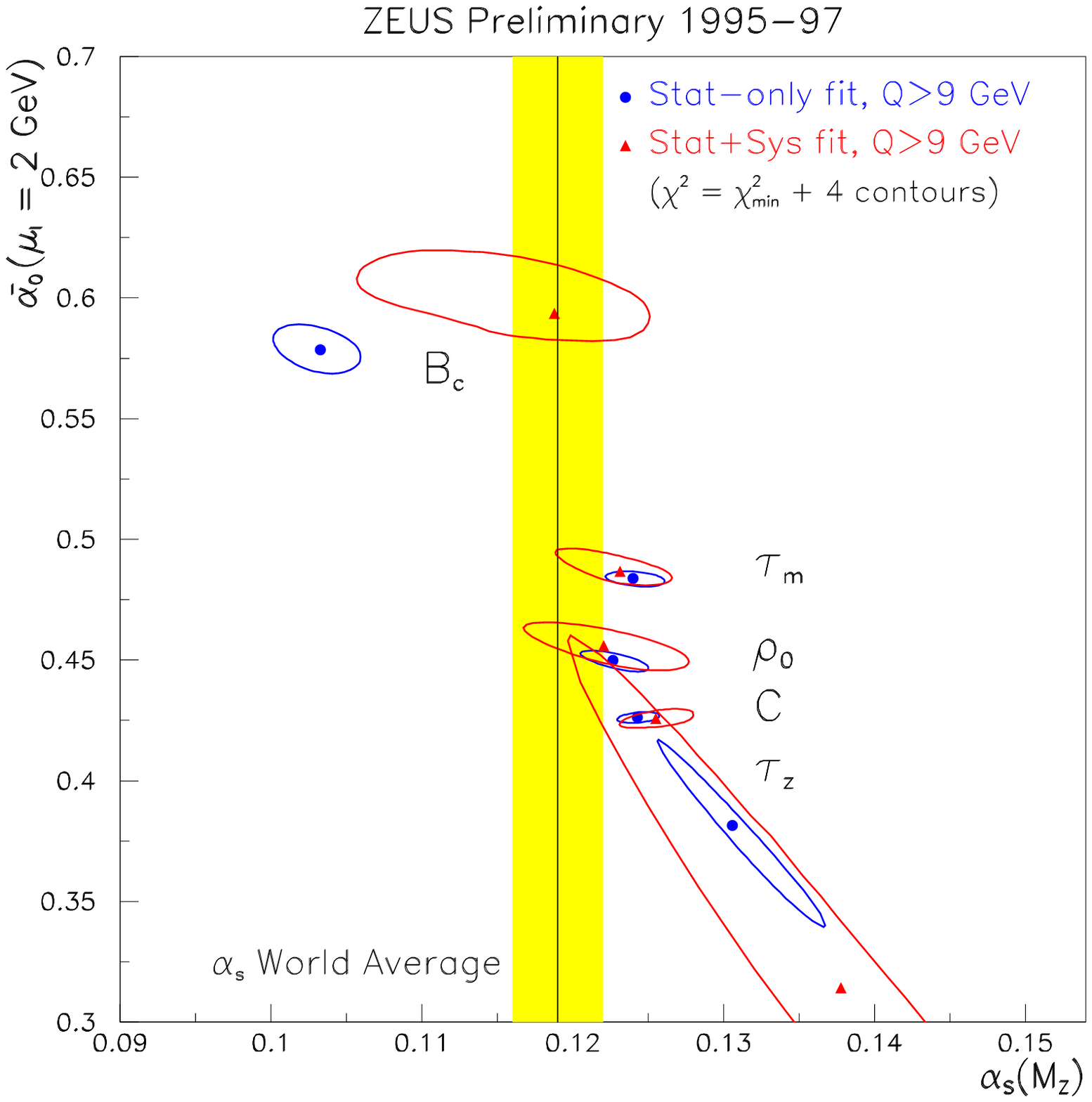,width=.32\textwidth} 
  \caption{Correlations of $\bar{\alpha}_0$ vs $\alpha_s(M_Z)$
    from power correction fits to {\sc Zeus} event shape means.
    Small contours include statistical errors only}
  \label{fig1} 
\end{figure}   
\begin{figure} \centering
  \epsfig{file=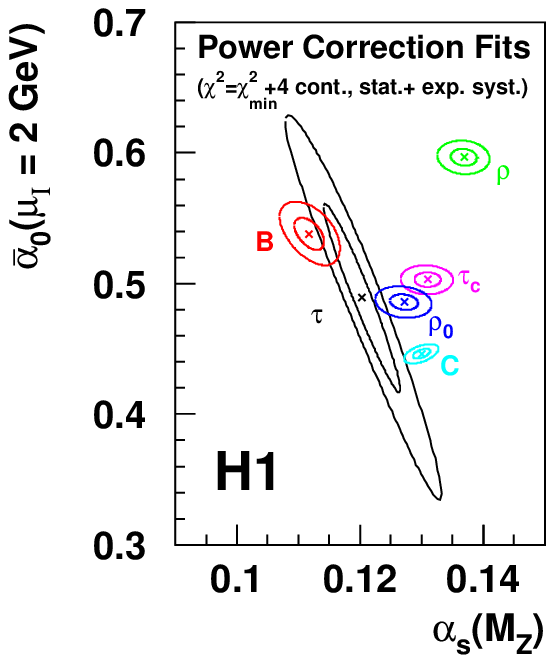,width=.45\textwidth} 
  \caption{Correlations of $\bar{\alpha}_0$ vs $\alpha_s(M_Z)$
    from power correction fits to {\sc H1} event shape means} 
  \label{fig1a}
\end{figure}

Hadronisation is treated within the concept of power corrections\cite{dw} with
a calculable coefficient $a_F$ and a universal function
    \begin{eqnarray}
      {\cal P} & = & 1.61\,\frac{\mu_I}{Q}\, 
      \left [ \, \bar{\alpha}_{0}(\mu_I) - \alpha_s(Q)
        \phantom{\frac{1}{2}}
      \right. \nonumber \\
      & & \left. 
        - 1.22\,\left (\ln\frac{Q}{\mu_I} + 1.45 \right ) 
        \alpha^2_s(Q)\,\right ] \, .
      \label{eq2}
    \end{eqnarray}
One expects a $1/Q$ behaviour (except for $y_{kt}$), which is multiplied
by terms involving the strong coupling $\as$ and a non-perturbative 
effective coupling $\an(\mi)$, defined at an infrared matching scale 
$\mi = 2~\GeV$.

\begin{figure}
  \centering   
  \epsfig{file=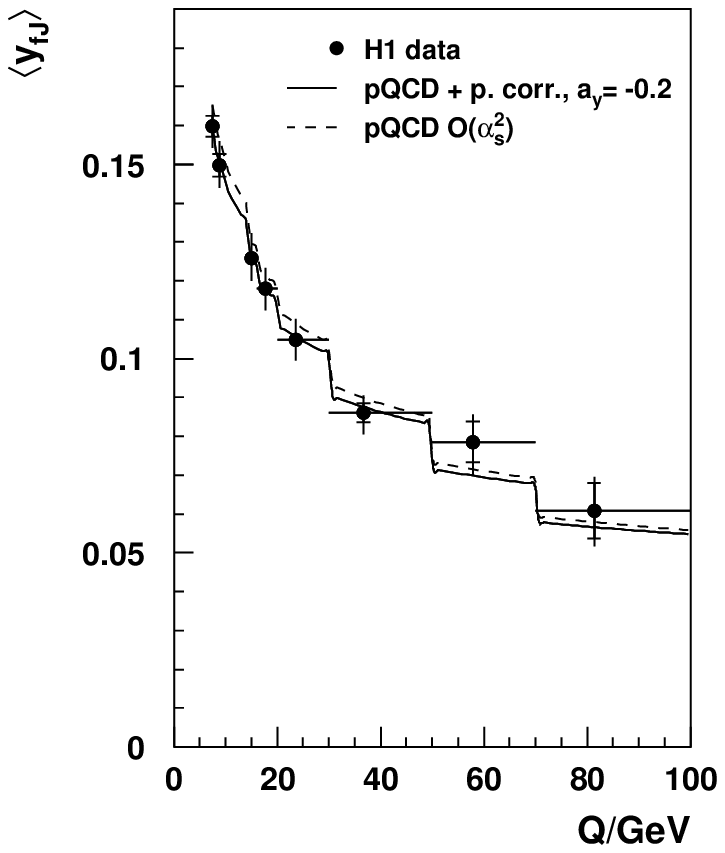,width=.32\textwidth} 
  \epsfig{file=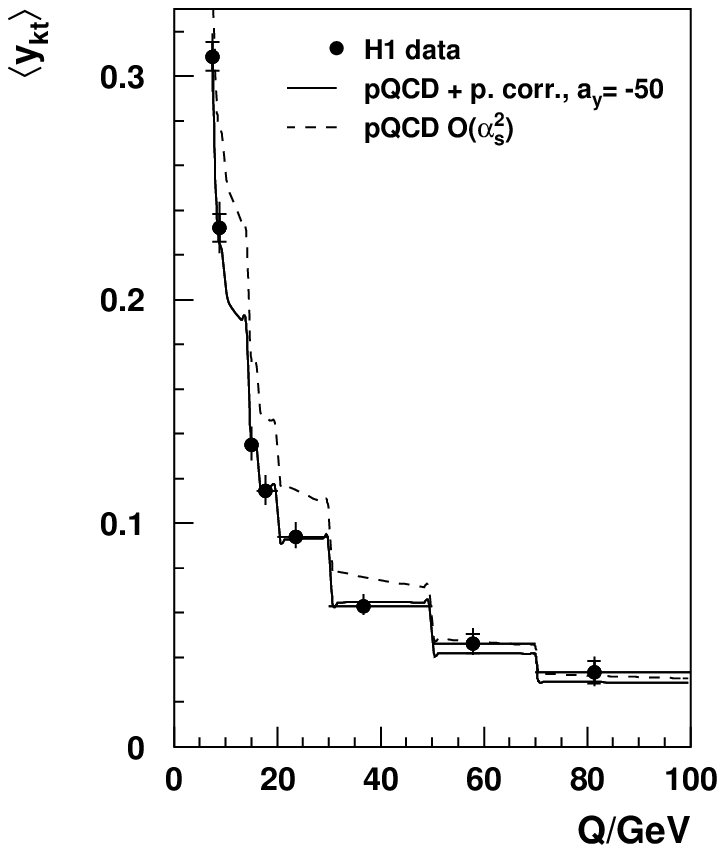,width=.32\textwidth}
  \caption{Mean values of two-jet rates $y_{fJ}$ and $y_{kt}$
    vs $Q$ of the H1 experiment}
  \label{fig2}
\end{figure}

The $Q$ dependences of the event shape means are well described by this ansatz.
Results of fits to $\an$ and $\asmz$ are shown as correlations 
in figs.~\ref{fig1} and \ref{fig1a};
the renormalisation scale uncertainties (not shown) exceed the experimental
errors by far.
For the non-perturbative parameter one finds a universal value of 
$\an \simeq 0.5 \pm 20\%$ for most observables.
{\sc Zeus} reports that fits to $\mean{B}$ and $\mean{\tau_z}$, 
resulting in a large spread (see fig.~\ref{fig1}),
are especially sensitive to experimental systematics 
and exhibit a significant $x$-dependence at low $Q$. 
In general, power corrections do not dependent on $x$;
such terms may, however, arise for $\mean{B}$. 
The H1 analysis, presented in fig.\ref{fig1a}, demonstrates
the strong influence of the treatment of hadrons on the jet masses 
$\rho$ and $\rho_0$.
A correction to massless hadrons, $\rho_0$, leads to a more consistent
interpretation of power corrections.
The spread of $\asmz$ is considerable for both experiments,
suggesting that higher order QCD contributions are missing. 
This is supported by large 
scale uncertainties.

The energy dependence of the H1 mean two-jet rates are shown 
in fig.~\ref{fig2}.
They exhibit much smaller hadronisation corrections than the other variables.
For the {\sc Jade} algorithm $\mean{y_{fJ}}$ the conjectured coefficient\cite{dw}
$a_{fJ}=1$ leads to an unphysically low value of $\an$ and is excluded by the data. 
Instead a small negative hadronisation contribution is preferred. 
In case of the $k_t$ algorithm no firm power correction prediction exists
except of a $1/Q^2$ dependence for $\mean{y_{kt}}$, very different from the 
other event shapes. 
Such a behaviour is supported by the H1 data, a $1/Q$ shape can be ruled out.
An experimental determination of the unknown parameters,
$a_{kt}$ and $\ao$ together with $\as$, 
suffers from large correlations\cite{h1}.
In view of the existing data more theoretical work on the jet rates 
is needed.

\begin{figure*}
  \centering    \mbox{
  \epsfig{file=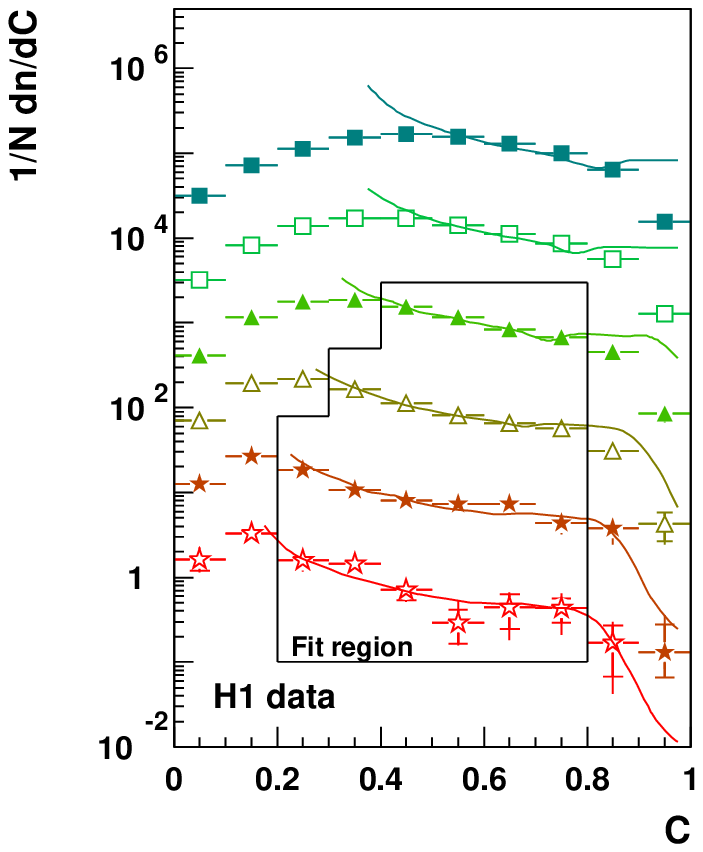,height=6.cm} 
  \epsfig{file=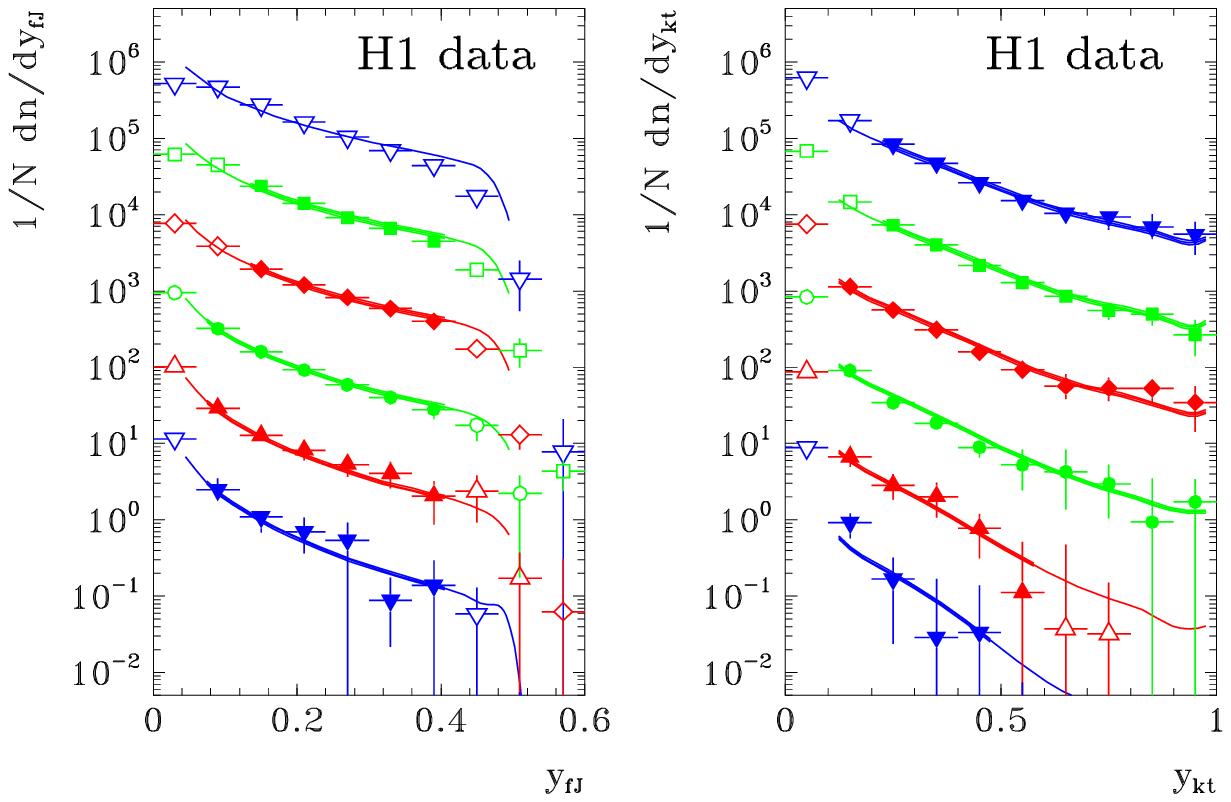,height=5.6cm} }
  \caption{Fits to differential distributions of 
    the $C$ parameter including power corrections 
    and the two-jet rates $y_{fJ}$ and $y_{kt}$ applying pQCD
    without hadronisation corrections. The H1 data cover a
    $\mean{Q}$ range from 15~GeV (top) to 81.3~GeV (bottom).}
  \label{fig3}
\end{figure*}

\section{Power corrections to spectra}

Power corrections to event shape spectra lead to a shift of the
pQCD prediction\cite{dw2}
\begin{equation}
  \frac{1}{\sigma_{{\rm tot}}}\,\frac{d\sigma(F)}{dF} = 
  \frac{1}{\sigma_{{\rm tot}}}\,
  \frac{d\sigma^{\rm{pert}}(F - a_F {\cal P})}{dF} \, , 
  \label{eq3}
\end{equation}
provided $\mu_I/Q < F < F_{max}$. 
The shift $a_F {\cal P}$ amounts to exactly the same value as for the mean values.
The event shape distributions at $Q>14~\GeV$ can be well described 
by eq.~(\ref{eq3}) within restricted regions, as shown in fig.~\ref{fig3}.
However, the fit values of $\an$ and $\asmz$ are, in general,
inconsistent (larger) to those from fits to the means. 
For example $(\an,\ \asmz) = (0.45,\ 0.130)$ from $\mean{C}$ and
$(0.62,\ 0.131)$ from $d\sigma/dC$. 
It is hoped that resummed QCD calculations\cite{dasgupta} will improve the 
applicability of power corrections to DIS event shape spectra.

The analysis of mean values lead to small  hadronisation corrections for the 
two-jet rates.
In fact, at sufficiently high energies $Q$, the jet rate spectra can be 
reasonably well described by pQCD alone, {\em i.e.} neglecting power corrections 
or hadronisation contributions completely.
This is shown in fig.~\ref{fig3} for $d\sigma/dy_{fJ}$ and $d\sigma/dy_{kt}$, 
using 0.116 and 0.118, respectively, for the strong coupling constant.

\section*{Summary}
Event shape studies of deep inelastic scattering provide very useful information
to get a better understanding of the interplay between perturbative and 
non-perturbative QCD. 
The basic concept of approximate universal power corrections
is generally supported by the {\sc Hera} experiments,
yielding a common  parameter $\an\simeq 0.5 \pm 20\%$. 
However, there remain several open questions. 
The quality of the data requires further theoretical progress concerning the jet 
rates and $x$-dependence of power corrections and resummed QCD calculations.

\end{document}